\documentclass[a4paper, onecolumn,11pt, floatfix,  uperscriptaddress, accepted=2023-03-07]{quantumarticle}
\pdfoutput=1
\usepackage[numbers,sort&compress]{natbib}
\usepackage[pdftex]{graphicx}
\usepackage[dvipsnames]{xcolor}
\usepackage{color}
\usepackage{amsmath}
\usepackage{amssymb}
\usepackage{dcolumn}
\usepackage{bbold}
\usepackage{braket}
\usepackage{lipsum} 
\usepackage{caption}
\usepackage{subcaption}
\usepackage{amsthm}
\usepackage{setspace}
\usepackage{hyperref}
\usepackage[T1]{fontenc}
\definecolor{darkblue}{RGB}{0,0,149}
\hypersetup{
    colorlinks=true,
    linkcolor=red,
    filecolor=darkblue,
    urlcolor=darkblue,
    citecolor=Green,
    linktocpage=true,
}
\usepackage{url}

\begin{document}
\newcommand{\beq}{\begin{equation}}
\newcommand{\eeq}{\end{equation}}

\title{Quantum States of Fields for Quantum  Split Sources}

\author{Lin-Qing Chen}%
 \email{linqing.nehc@gmail.com}
\affiliation{Centre for Quantum Information and Communication,
Université Libre de Bruxelles, 1050 Brussels, Belgium.}
\affiliation{Institute for Theoretical Physics, ETH Z{\"u}rich, 8093 Z{\"u}rich, Switzerland}

\author{Flaminia Giacomini}%
 \email{fgiacomini@phys.ethz.ch}
\affiliation{Institute for Theoretical Physics, ETH Z{\"u}rich, 8093 Z{\"u}rich, Switzerland}
\affiliation{Perimeter Institute for Theoretical Physics, 31 Caroline St. N, Waterloo, Ontario, N2L 2Y5, Canada.}

\author{Carlo Rovelli}%
\email{rovelli.carlo@gmail.com}
\affiliation{Perimeter Institute for Theoretical Physics, 31 Caroline St. N, Waterloo, Ontario, N2L 2Y5, Canada.}
\affiliation{Aix-Marseille University, Universit\'e de Toulon, CPT-CNRS, F-13288 Marseille, France.} 
\affiliation{Department of Philosophy and the Rotman Institute of Philosophy, 1151 Richmond St.~N London  N6A5B7, Canada.}

%\date{\today}

\begin{abstract}

\noindent 
{\bf Abstract:} 
Field mediated entanglement experiments probe the quantum superposition of macroscopically distinct field configurations. We show that this phenomenon can be described by using a transparent quantum field theoretical formulation of electromagnetism and gravity in the field basis. The strength of such a description is that it explicitly displays the superposition of macroscopically distinct states of the field. In the case of (linearised) quantum  general relativity, this formulation exhibits the quantum superposition of geometries giving rise to the effect.

\end{abstract}

\maketitle

%\tableofcontents
\section{Introduction}

The prospect of detecting the  entanglement induced by a  gravitational interaction is opening a new perspective on the possibility of observing phenomena at the interface between quantum theory and gravity~\cite{Bose2017, marletto2017gravitationally, hall2018two, anastopoulos2018comment, belenchia2018quantum, belenchia2019information, christodoulou2019possibility, howl2020testing, marshman2020locality, krisnanda2020observable, Marletto:2020cdx, Galley:2020qsf, pal2021experimental, Carney:2021yfw, kent2021testing,   danielson2021gravitationally,   Zhou:2022frl, Danielson:2022tdw, Bose:2022uxe, Polino:2022qxg}. Field Mediated Entanglement (FME), indeed, could provide evidence for the existence of quantum superpositions of gravitational fields~\cite{cecile2011role,zeh2011feynman,Blencowe:2012mp,Anastopoulos:2013zya,Anastopoulos:2015zta,carlesso2017cavendish,bahrami2015gravity,ford1982gravitational,kafri2013noise,kafri2014classical,altamirano2018gravity,anastopoulos2020quantum,Anastopoulos:2021egs}.
%The idea of testing the quantum superposition of gravitational fields~\cite{cecile2011role,zeh2011feynman,Blencowe:2012mp,Anastopoulos:2013zya,anastopoulos2015probing,carlesso2017cavendish,bahrami2015gravity,ford1982gravitational,kafri2013noise,kafri2014classical,altamirano2018gravity,anastopoulos2020quantum,Anastopoulos:2021egs} has recently raised a large interest \fedit{thanks to the realization that Field Mediated Entanglement (FME) could provide an experimental evidence of this phenomenon~\cite{Bose2017, marletto2017gravitationally, hall2018two, anastopoulos2018comment, belenchia2018quantum, belenchia2019information, christodoulou2019possibility, howl2020testing, marshman2020locality, krisnanda2020observable, Marletto:2020cdx, galley2020no, pal2021experimental, Carney:2021yfw, kent2021testing,   danielson2021gravitationally,   Zhou:2022frl, Danielson:2022tdw, Bose:2022uxe, Polino:2022qxg}.} 
The experimental verification of this effect might be achieved in the foreseeable future, thanks to the impressive experimental progress both in tests of the gravitational field associated to lighter and lighter particles~\cite{westphal2020measurement}, ground state cooling~\cite{Delic:2020ndp, Magrini:2020agy, tebbenjohanns2021quantum}, and generation of large superpositions~\cite{fein2019quantum, kovachy2015quantum, Overstreet:2021hea}. Hence, it is now it possible to devise a new generation of experiments which will eventually test the gravitational field associated to a quantum split source~\cite{Aspelmeyer:2022fgc}. These new developments have raised the hope to open the first phenomenological window on a detectable quantum gravitational phenomenon. 

While the intuitive physics underpinning FME is simple, its detailed field theoretical description is not, because it involves a quantum superposition of macroscopically distinct field configurations. In the conventional Fock basis description, each gravitational field configuration contains an infinite number of particles. On the other hand, the interacting potential can be described in quantum field theory as an exchange of virtual particles.
%In this regime, each gravitational field configuration contains an infinite number of particles in the conventional Fock basis description, \LQ{and the Coulomb/Newtonian potentials are mathematically described as exchanging virtual particles between sources in the quantum field theory}\fla{This might be a bit too technical for the introduction} \LQ{hmm, it is textbook QFT, plus  the virtual particle was many people's confusion, hence it motivated us well?}.
These aspects of the formulation have raised  questions regarding the exact physical implications of a potential detection of a gravitational FME. These include the role of the Newtonian potential and the split between transverse and longitudinal components of the gravitational field~\cite{anastopoulos2018comment}, the relation between the quantum gravitational state and the superposition of geometries, the precise mechanism leading to the mediation of entanglement between the source systems~\cite{hall2018two}, and the role of locality in mediating the interaction~\cite{Christodoulou:2022knr, Fragkos:2022tbm}.

 Here, we give a straightforward and transparent Hilbert space description of the phenomenon, and in particular of the quantum state of the gravitational field sourced by a quantum particle in a spatial superposition. This description sheds light on the open questions mentioned above (a field theoretical account of the relative phase has been given in \cite{christodoulou2022locally} using path integrals).  Specifically, we use the Schr{\"o}dinger representation of quantum fields in the field basis~\cite{hatfield2018quantum} and we apply it to an effective quantum field-theoretic description of gravity in the weak regime~\cite{Kuchar:1970mu, Burgess:2003jk, Donoghue:1994dn, Donoghue:2017ovt, bern2002perturbative}. For the vacuum state, we reproduce certain results in~\cite{Kuchar:1970mu}, which adopted a different approach.  The field basis allows us to treat naturally quantum split macroscopic sources. We fix the gauge minimally to emphasise the gauge-independent aspects of the description. In the electromagnetic case, we reproduce features of the Dirac dressing formalism~\cite{Dirac2011} as well as related result by the BRST quantization~\cite{Barnich:2010bu}.

We start by illustrating the general structure of the effect in an elementary finite dimensional toy model (Section~\ref{sec:ToyModel}), then we  analyse both the electromagnetic (Section~\ref{sec:EM}) and the gravitational (Section~\ref{sec:LinQG}) cases separately. For the latter, we give a simple derivation of the hamiltonian formulation of linearized gravity and provide concise steps of its quantization.  We rigorously derive the quantum state of the gravitational field corresponding to the Newtonian field of a superposition of  masses in the time gauge. This proves that the Newtonian field has a quantum state which is entangled with the source state, as previously argued~\cite{belenchia2018quantum,belenchia2019information}. Finally, the relative phase accumulated in the interferometer when the Hamiltonian acts on this joint quantum state of gravity and matter coincides with the phase predicted when the two particles interact via the (nonlocal) Newtonian potential. In Section~\ref{sec:Discussion} we discuss the physical interpretation of our results.  

Greek indices run over all spacetime components, Latin indices indicate the spatial components.

\begin{figure}[t]
\centering
\includegraphics[width=0.5\textwidth]{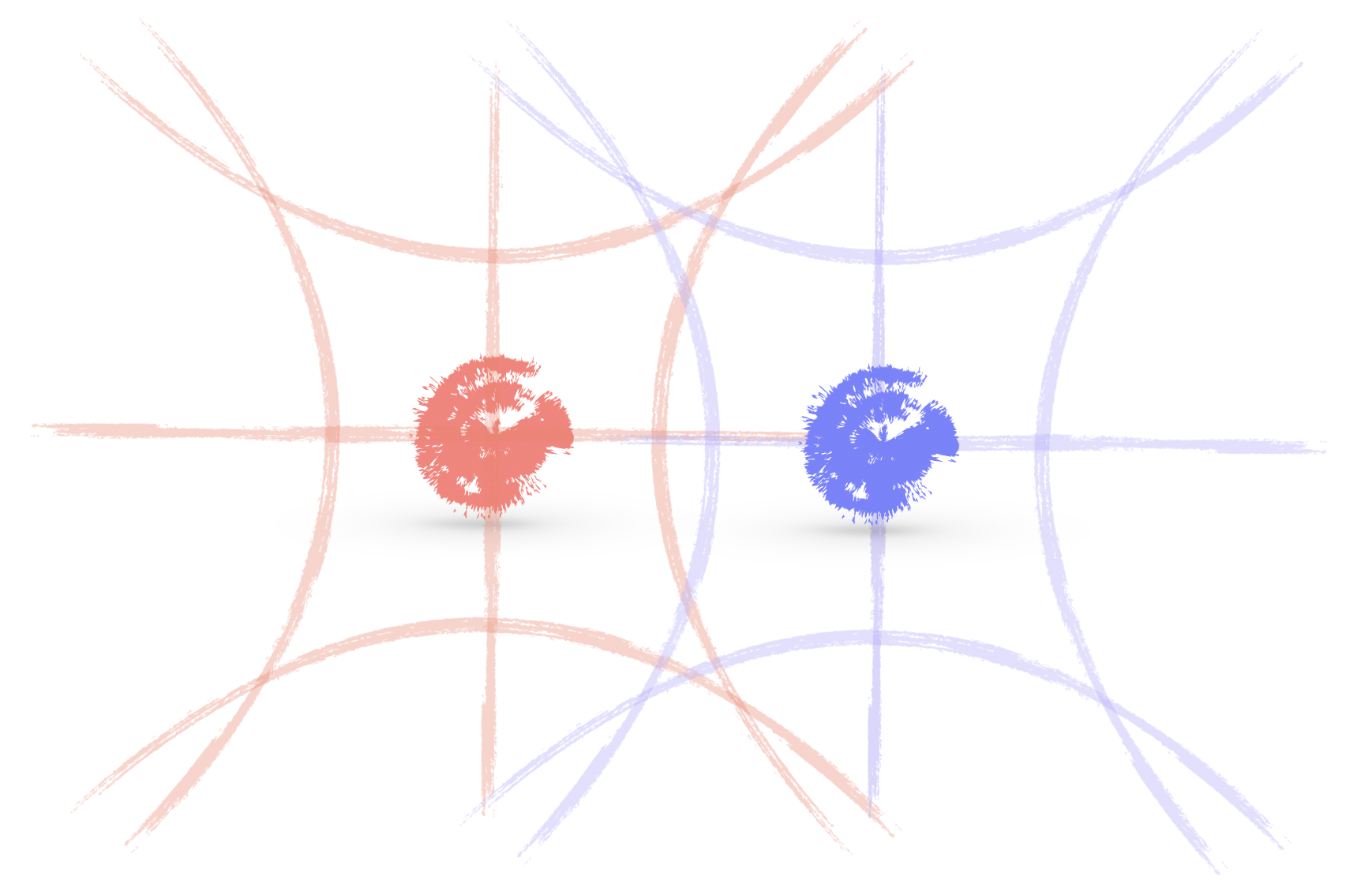}
\caption{A pictorial representation of the superposition of two locations of a massive particle, and the corresponding superposition of the gravitational field it generates.}
\end{figure}

\section{Toy model}
\label{sec:ToyModel}

Consider a one dimensional  harmonic oscillator (HO) with mass $m$ and spring constant $k$ affected by an external force generated by a linear potential. The Hamiltonian that describes its dynamics is
\begin{equation}
	H_\gamma = \frac{p^2}{2m} + \frac{k}{2}x^2 - m\gamma x,
\end{equation}
where $\gamma$ determines the strength of the force. Notice that $\gamma$ appears as a source term in the equation of motion
\begin{equation}
	\ddot{x} + \omega^2 x = \gamma,
\end{equation}
where $\omega^2 = k/m$. The effect of the source is to add a constant term to the 
general solution of a free harmonic oscillator:
\begin{equation}
	x(t) = A \sin(\omega t + \phi_0) + \frac{\gamma}{\omega^2}\equiv A \sin(\omega t + \phi_0) + x_\gamma
\end{equation}
Notice that the presence of the source shifts the minimum-energy classical solution of the free theory $x(t)=0$ to the solution
\begin{equation}
x(t)= x_\gamma.
\end{equation}
The effect of the source, in fact, is simply to displace the minimum of the potential. 

The quantum theory can be written in the Schr\"odinger representation, in terms of wave functions $\psi(x,t)$. The spectrum of the Hamiltonian 
is easy to find. Writing $\tilde  x =x-x_\gamma$ puts the Hamiltonian in the form  
\begin{equation}
	\hat{H} = \frac{{p}^2}{2m} + \frac{k}{2}\tilde x^2 -\frac{m\gamma^2}{2\omega^2},
\end{equation}
which is the free HO Hamiltonian minus a constant term. Therefore, the spectrum of the energy is the HO spectrum, but shifted by the constant term. 

Of particular interest to us is the minimal energy state, or "vacuum". Clearly, this is the standard HO vacuum shifted to $x_\gamma$:
\begin{equation}
	\psi_\gamma(x)=\psi_o(x-x_\gamma)=
	C e^{-\frac{(x-x_\gamma)^2}{2\sigma^2}}
\end{equation}
with $\sigma^2=\hbar/(m\omega)$. In the momentum representation, the shift in $x$ becomes a phase:
\begin{equation}
	\psi_\gamma(p)=\psi_o(p)\, e^{-i\hbar x_\gamma p}=
	C e^{-\frac{\sigma^2 p^2}{2}}\, e^{-\frac{i}\hbar x_\gamma p}.
\end{equation}
Formally, we can introduce the translation operator \begin{equation}
T_\gamma = e^{-\frac{i}{\hbar}x_\gamma {p}}  
\label{T}
\end{equation}
and write
\begin{equation}
	\psi_\gamma =T_\gamma\, \psi_o.
\end{equation}
We have illustrated this with a certain detail, because a very similar structure appears in the quantum states of the field in the presence of sources. 

Consider the case in which $\gamma$ is a parameter that we can manipulate at will. If the HO is in the minimum energy state  $\psi_\gamma$ and we change $\gamma$ to $\gamma'$, the movement of the source excites the HO and moves it away from the minimum energy state. If there is also a dissipation sink into which the HO can dissipate energy, after a transient phase the HO re-settles to the minimum energy state corresponding to the new value of $\gamma'$, namely to $\psi_{\gamma'}$.  Keeping in mind the existence of transient phase is important for the following.  

Next, let us now consider a situation where the constant $\gamma$ is replaced by a quantum physical variable $\gamma$.  In this case, the quantum states of the coupled system can be represented in the Schr\"odinger representation as wave functions $\Upsilon(x,\gamma)$.
For simplicity, let us  assume that the dynamics of $\gamma$ is trivial and its mass is very large. (The only quantum aspect relevant for what follows is that the variable can be set into a quantum superposition of different values.) This mimics the physical regime in which we will be interested for the FME case. In this case, semiclassical states corresponding to classical configurations where $\gamma$ takes a specific value $\gamma=\gamma_i$ are well approximated by the  generalised eigenstates $\phi_{\gamma_i}(\gamma)\sim \delta(\gamma-\gamma_i)$~\cite{chevalier2020witnessing}. In the approximation considered, these are stationary. The (approximately)  stationary state of the entire system where $\gamma$ takes a specific value $\gamma=\gamma_i$ which has minimal energy is therefore 
\begin{equation}
	\Upsilon_{\gamma_i}(x,\gamma) =\phi_{\gamma_i}(\gamma)\,  \psi_{\gamma_i}(x).
\end{equation}
The states that will interest us below are states where the variable $\gamma$ is in the superposition of n values: $i=1,...n$. If the variable gamma is set to such a superposition, the stationary minimum energy state for the entire system will be
\begin{equation}
	\Upsilon(x,\gamma) = \sum_i c_i\  \phi_{\gamma_i}(\gamma)\, \psi_{\gamma_i}(x)
\end{equation}
where the $x$ and $\gamma$ variables are entangled. 

We are now going to write analog states for field theories. The variable $\gamma$ is the analog of the quantum sources set in superposition, and the variable $x$ is the analog of the field variables.   We will emphasize similarities and differences between this simple case and the field case, both in electromagnetism and in gravity.

\section{Electromagnetic field in the Schr\"odinger representation}
\label{sec:EM}

\subsection{Free theory with no sources}
The action of the free electromagnetic field is
\begin{equation}
	S = -\frac{1}{4} \int d^4 x\ F_{\mu\nu}(x)F^{\mu\nu}(x),
\end{equation}
with $F_{\mu\nu}(x) = \partial_\mu A_\nu -  \partial_\nu A_\mu$. The equations of motion are
\begin{equation}
	\square A^\mu - \partial^\mu (\partial_\alpha A^\alpha) = 0.
\end{equation}
The action does not depend on $\partial_0 A_0 (x)$. This  implies a primary constraint:  the momentum canonically conjugated  to $A_0(x)$, vanishes. We can set $A_0(x)=0$ strongly. The other momenta are
\begin{equation}
	\pi_i = \frac{\partial \mathcal{L}}{\partial \dot{A}_i} = \partial_0 A_i -  \partial_i A_0 = E_i,
\end{equation}
 with $E_i$ being the components of the electric field. The secondary constraint $\cal G$ obtained from the commutator of the Hamiltonian with the primary constraint is the Gauss law
\begin{equation}
{\cal G} = 	\nabla \cdot  E = 0.
\label{G}
\end{equation}
In the Hamiltonian framework, the theory can be defined by  the Poisson bracket structure \begin{equation}
	\left\{ A_i(\vec{x}, t), E_j(\vec{x}\,', t) \right\} = \delta_{ij} \delta^{(3)} (\vec{x}- \vec{x}\,'),
\end{equation}
the Gauss law constraint \eqref{G} and the Hamiltonian \begin{equation} \label{eq:HamEM}
	{H} = \frac{1}{2}\int d^3 x \ ( E^2(\vec x) + B^2(\vec x))
\end{equation}
where $B=\nabla \times A$ is the magnetic field. Since we have set $A_0=0$, there are no constraint terms to add to the Hamiltonian. 

The quantization of the electromagnetic field in the Schr{\"o}dinger representation then can be carried out in terms of wave functionals $\Psi[A]$ on which the $A$ field operator acts diagonally and the $E$ field operator acts as 
\begin{equation}
 E_i(\vec{x})  = -i \hbar  \frac{\delta}{\delta  A_i(\vec{x})}.
\end{equation}
(from now on, we set $\hbar=1$.) This is the "field basis" representation of quantum  electromagnetism.  The Gauss constraint can be imposed strongly on the states.  That is, the wave functionals are restricted to those satisfying 
${\cal G}\Psi=0$, namely
\begin{equation}
 \partial_i \frac{\delta}{\delta A_i(\vec{x})} \Psi[A] =0.
 \label{qG}
\end{equation}
It is easy to see that the operator $\cal G$ is the generator of the gauge transformation $A\to A=d\lambda$, hence the general solution of Eq.~\eqref{qG} is given by the functionals that are constant along the gauge orbits. That is
\begin{equation}
 \Psi[A]= \Psi[B[A]],
\end{equation}
since the magnetic field is gauge invariant and gauge inequivalent $A$'s have different magnetic fields. 

Like for the Harmonic oscillator, the minimum energy state of the Hamiltonian is a Gaussian centered on the classical minimum energy configuration $B=0$. Explicitly, it is 
\begin{equation}
 \Psi_o[A]= C \exp\left\{-\frac{1}{2 \hbar}\int d^3 x d^3 y \ \frac{B(\vec x) \cdot B(\vec y)}{|\vec x-\vec y|^2}\right\}.
 \label{vEM}
\end{equation}
which is the gauge invariant  ground state functional of the
electromagnetic field~\cite{WheelerG, hatfield2018quantum}.
 This is the quantum state that corresponds to the Fock vacuum $|0\rangle$ of the standard Fock formulation of quantum field theory in this basis: $\Psi_o[A]=\langle A| 0\rangle$. Notice that the exponent is non local in the field: this is a reflex of the well known non-local correlations in the vacuum of quantum field theory. 

It is convenient for the following to express the field in Fourier transform. Defining 
%imposing the equal-time commutation relations
%\begin{equation}
%	\left[ A_i(\vec{x}, t), E_j(\vec{x}\,', t) \right] = i \delta_{ij} \delta^{(3)} (\vec{x}- \vec{x}\,'),
%\end{equation}
%and in a basis where
%\begin{equation}
%A_i(\vec{x}) \ket{a} = a_i(\vec{x})\ket{a}, \qquad E_i(\vec{x}) \ket{a} = i \frac{\delta}{\delta  a_i(\vec{x})}\ket{a}.
%\end{equation}
%\begin{equation} \label{eq:HamEM}
%	\hat{H} = \frac{1}{2}\int d^3 x \left\lbrace - \frac{\delta}{\delta a_i(\vec{x})} \frac{\delta}{\delta a_i(\vec{x})} + (\nabla \times a(\vec{x}))\cdot (\nabla \times a(\vec{x})) \right\rbrace,
%\end{equation}
%and the quantum state $\Psi[a] = \braket{a|\Psi}$ satisfies the time-independent Schr{\"o}dinger equation $H \Psi[a] = E \Psi[a]$ and the Gauss constraint $\partial_i \frac{\delta}{\delta a_i(\vec{x})} \Psi[a] =0$.
%It is easier to find the vacuum state of the Hamiltonian  Fourier transforming the fields, namely writing 
\begin{equation}
	A_i (\vec{x}) = \int \frac{d^3 k}{(2\pi)^{3}} \tilde{A}_i (\vec{k}) e^{i \vec{k}\cdot \vec{x}}, \qquad
	 \frac{\delta}{\delta A_i(\vec{x})} = \int \frac{d^3 k}{(2\pi)^{3}} \frac{\delta}{\delta \tilde{A}_i(\vec{k})} e^{-i \vec{k}\cdot \vec{x}},
\end{equation}
The vacuum state can be written as 
\begin{equation} 
	\Psi_o[A]= C \exp \left\{ -\frac{1}{2 \hbar}\int \frac{d^3 k}{(2\pi)^3} \frac{(\vec{k} \times \tilde{A}(\vec{k}))\cdot (\vec{k} \times \tilde{A}(-\vec{k}))}{|\vec{k}|}\right\}.
\end{equation}

% The ground state of the Hamiltonian $\Psi_0[a]$ can be found using the Ansatz
%\begin{equation}
 %\Psi_0[a]= \eta \exp \left[- G_F (\tilde{a}) \right],
%\end{equation}
%where $\eta$ is a normalisation constant and 
Notice that rearranging the terms using the vector identity $(\vec{a}\times \vec{b}) \cdot (\vec{c}\times \vec{d}) = (\vec{a}\cdot \vec{c}) (\vec{b}\cdot \vec{d}) - (\vec{a}\cdot \vec{d})(\vec{b}\cdot \vec{c})$ this can be rewritten in the form 
\begin{equation} \label{eq:VacuumEMFree}
	\Psi_o[A] = C \exp\left\{-\frac{1}{2\hbar}\int \frac{d^3 k}{(2\pi)^3} |\vec{k}| \left(\tilde{A}_i(\vec{k})  P^i_j(\vec{k}) 
 \tilde{A}^j(-\vec{k}) \right)\right\}.
\end{equation}
where $P^i_j(\vec{k})$ is the projector onto the transverse direction of the momenta
\beq
P^i_j(\vec{k}) = \delta^i_j -\frac{k^i k_j}{|\vec{k}|^2}.
\label{P}
\eeq
We shall also use the projector on the longitudinal direction
\beq
\Pi^i_j(\vec{k}) = \frac{k^i k_j}{|\vec{k}|^2}.
\label{Pi}
\eeq
The definition of these  operators requires care for $|\vec{k}|^2 =0$. In coordinate space, $1/|\vec{k}|^2$ is the inverse of the Laplacian, therefore the sector of the fields where the definition is ambiguous is given by the harmonic functions. We take these to be longitudinal by definition. 

The Gauss constraint, indeed, states that $\Psi[A]$ is constant in the direction of the longitudinal components of the field. 

Recall that any vector field $A$ can be split into a longitudinal component which can be written as $A^L=\nabla f$, and a transverse component $A^T$ that satisfies $\nabla\cdot A^T=0$. The decomposition is not unique, because if $\lambda$ is harmonic ($\Delta \lambda=0$) then $A=\nabla \lambda$ satisfies $\nabla\cdot A=0$, so it is both longitudinal and transverse. If boundary conditions are set the harmonic part of the field is uniquely determined by these. The split between a longitudinal and a transverse component of the field is therefore non-local.  

Let us also illustrate the theory in the representation that diagonalizes the electric field $E$ rather than the potential $A$. This is easily obtained by a functional Fourier transform in field space from the wave functionals $\Psi[A]$ of Eq.~\eqref{eq:VacuumEMFree} to wave functionals $\Psi[E]$. (This is analog to going from the $\psi(x)$ to the $\psi(p)$ representation in the toy model.)  The Gauss law is diagonal in this representation and therefore its solution is given by the states that have support on the fields $E$ satisfying the Gauss law. (The Fourier transform of a constant is a delta function.) That is, all the physical states of the theory have the form 
\begin{equation}
    \Psi[E]=\delta[\nabla \cdot  E]\ \Phi[E].
\end{equation}
The vacuum is easily obtained in the same manner. It can be written in the form
\begin{equation} \label{eq:AnsatzEMFree}
	\Psi_o[E] = C\ \delta[\nabla \cdot  E]\  
	 \exp\left\{-\frac{1}{2 \hbar}\int d^3 x d^3 y \ \ \frac{E(\vec x) \cdot E (\vec y)}{|\vec x-\vec y|^2}\right\}.
\end{equation}
Importantly for the following, notice that the state is not independent of the longitudinal components of the field: it is peaked around their vanishing value. 

\subsection{Electromagnetism with a source}\label{EWaS}

The action of the electromagnetic field theory with a source, whose 4-current is $J_\mu (x)$, is
\begin{equation}
	S = -\frac{1}{4} \int d^4 x\  F_{\mu\nu}(x)F^{\mu\nu}(x) - \int d^4 x\  J_\mu(x)A^\mu(x).
\end{equation}
The equations of motion are
\begin{equation}
	\square A^\mu - \partial^\mu (\partial_\alpha A^\alpha) = - J^\mu(x).
\end{equation}
To start with, we consider the source to be external, fixed and static. Then it has only the component $J^0(x) \equiv \rho(x)$, which is the charge density. 
%Analogously to the case without sources, the Lagrangian does not depend on $\partial_0 A_0(x)$, so we have $\pi_0(x) =0$. We hence fix $A^0(x) =0$ and obtain from the $\mu=0$ equation of motion
The hamiltonian theory is the same as above, with the only difference that the Gauss law constraint is now
\begin{equation}
    {\cal G}_\rho = \nabla\cdot E - \rho = 0
\end{equation}

Before going to the quantum theory, let us pause a moment to consider a minimum energy classical solution in this formalism. Say the source is concentrated in the origin:
\begin{equation}
    \rho(\vec x)\sim  q \delta(\vec x). 
\end{equation}
The equations of motion have of course many possible solutions; let's focus on the minimum energy one.  The electric field is the Coulomb field
\begin{equation}
    E_q(\vec x)=-q\,\nabla \frac{1}{r}.
\end{equation}
where $r=|\vec x|$, and the magnetic field vanishes $B_e=0$.   But what about the potential $A$? A moment of reflection shows that the potential of the Coulomb field of a point-like charge {\em in this gauge} must be the {\em time dependent} field 
\begin{equation}
    A_q(\vec x, t)=-q\, t\; \nabla \frac{1}{r},
\end{equation}
because this gives the correct electric and magnetic fields in the time gauge $A_0=0$. It may be surprising at first that that the Coulomb field is time dependent in the time gauge, but this is the case.  As we shall see, the same is true in gravity for the Newtonian field. For a generic source $\rho$ 
\begin{equation}
    E_\rho(\vec x)=-\int d^3 y \ \nabla_x\cdot  \frac{\rho(\vec y)}{|\vec x-\vec y|}, \qquad
    A_\rho(\vec x)= -t \int d^3 y \ \nabla_x\cdot  \frac{\rho(\vec y)}{|\vec x-\vec y|}. 
\end{equation}
 For an ensemble of pointlike charges $q_n$ located in the points $x_n$, the electric field is 
\begin{equation}
    E_{x_n}(\vec x)=-\sum_n  \nabla_x  \frac{q_n}{|\vec x-\vec x_n|}.
\end{equation}

Let us move to the quantum theory. The Hamiltonian is the same as before, but the space of the allowed states is different.  These must satisfy the modified Gauss law constraint ${\cal G}_\rho$. 
In the $\Psi[E]$ representation, this implies simply a modification of the delta function:
\begin{equation}
    \Psi[E]=\delta[\nabla \cdot  E-\rho]\ \Phi[E].
\end{equation}
In the $\Psi[A]$ representation, this implies that the state is not independent from the longitudinal components of $A$. Rather, its dependence on these is determined by $\rho$ via the Gauss constraint. In both representations, the minimal energy state can be obtained as we did in the toy model, by introducing the field space translation operator analogous to the operator \eqref{T}
\begin{equation}
T_\rho = e^{-\frac{i}{\hbar}\int d^3 x\  E_\rho(\vec x) A(\vec x)}  
\end{equation}
This shows that the minimal energy state can be written in the $\Psi[A]$ representation simply shifting Eq.~\eqref{vEM}, that is\footnote{Strictly speaking, the state \eqref{shifteddstate} belongs to a different Hilbert space than the state \eqref{vEM}, as the two correspond to different measures. These subtleties can in principle be addressed in an algebraic formulation considering the states as linear functionals on the observable algebra; but they are not relevant in what follows because we are ultimately interested in linear superpositions between states that have the same total charge, hence belong to the same Hilbert space.}
\begin{equation}
 \Psi_\rho[A]= C \exp\left\{-\frac{1}{2 \hbar}\int d^3 x d^3 y \ \frac{B(\vec x)\cdot B(\vec y)}{|\vec x-\vec y|^2} -\frac{i}{\hbar} \int d^3  x  E_\rho(\vec x)\cdot A(\vec x)\right\}. \label{shifteddstate}
\end{equation}
This linear shift is related to the Dirac dressing \cite{Dirac2011}. 

%We consider the limit in which the source moves very slowly, so that $J_i(x) \approx 0$. In this regime, we can neglect the coupling between the source and the field, meaning that the full Hamiltonian is still the one of Eq.~\eqref{eq:HamEM} and the only change compared to the previous section is that the Gauss law is now expressed in the basis of the fields over momenta as
%\begin{equation} \label{eq:GaussEMsource}
%	\left[k_i \frac{i \delta}{\delta \tilde{a}_i(k)} + \tilde{\rho}(k) \right]\Psi[\tilde{a}] =0.
%\end{equation}
%We now use a similar Ansatz to the source-free case, where
%\begin{equation}
% \Psi_0[a]= \eta \exp \left[- G_F (\tilde{a}) - G_I (\tilde{a})\right],
%\end{equation}
%where $\eta$ is a normalisation constant, $G_F (\tilde{a})$ is the same as in the previous section, and 
%\begin{equation} \label{eq:AnsatzEMFree}
%	G_I (\tilde{a}) = i\int d^3 k \tilde{f}(k) k^j \tilde{a}_j(k) \tilde{\rho}(k).
%\end{equation}
In terms of the Fourier components $\tilde A(\vec{k})$ this reads
%This Ansatz solves the modified Gauss constraint of Eq.~\eqref{eq:GaussEMsource} with $\tilde{f}(k) = |k|^{-2}$. Hence, the ground state of the electromagnetic field with a source is
\begin{equation}
\label{stateEMsource}
	\Psi_\rho[A]= C \exp \left[- \frac{1}{2 \hbar}\int \frac{d^3 k}{(2\pi)^3} \frac{(\vec{k} \times \tilde{A}(\vec{k}))\cdot (\vec{k} \times \tilde{A}(-\vec{k}))}{|\vec{k}|} + \frac{i}{\hbar}  \int \frac{d^3 k}{(2\pi)^3}  \frac{k^j \tilde{A}_j(\vec{k})}{|\vec{k}|^2} \tilde{\rho}(\vec{k}) \right].
\end{equation}
Notice that the last term depends on the \emph{longitudinal}  components of the field. In the $\Psi[E]$ representation, the translation operator gives
\begin{equation} \label{eq:AnsatzEMFree}
	\Psi_\rho[E] = C\ \delta[\nabla \cdot  E-\rho]\ \exp\left\{-\frac{1}{2 \hbar}\int d^3 x d^3 y \ \ \frac{(E(\vec x)-E_\rho(\vec x)(E(\vec y)-E_\rho(\vec y)}{|\vec x-\vec y|^2}\right\}.
\end{equation}

The modification of the minimal-energy eigenvalue due to the presence of the source is
\beq
{\cal E}_\rho - {\cal E}_o =  \int\frac{d^3k}{(2\pi)^3} \frac{\rho^2(\vec{k})}{|\vec{k}|^2}=  \int {d^3  x}\, d^3  y\  \frac{\rho(\vec{x})\rho(\vec{y})}{|\vec{x}-\vec{y}|}=\frac{1}{2}\int d^3  x\ (E_\rho^2+B_\rho^2)
\eeq
which is easily recognized as the potential energy stored in the field. 

%\CR{I have a doubt about the $\frac{1}{8\pi}$.   I think that this corresponds to (42). But it does not fit with the Hamiltonian... I do not know if this has to do with the fact that we do not have $4\pi$ in the Gauss law....  I am missing a constant...}

For a point-like  charge, the Fourier coefficient $\rho(\vec{k})$ of the delta function is simply 1. Notice that $1/|\vec{k}|^2$ is the Fourier coefficient of the Coulomb potential\footnote{Rigorously speaking, we need to Fourier transform $\mathcal{F} \left( e^{-a r}/r \right) = \frac{4 \pi}{k^2 + a^2}$ and then take $a\rightarrow 0$ limit. }. The self-energy of a pointlike particle 
\beq
{\cal E}_q - {\cal E}_o  =  \int d^3x\  \frac{q^2}{|x|}
\eeq
diverges. This is the usual self-energy classical divergence of point like charges and can be subtracted away.  Doing so, the energy of an ensemble of point like charges $q_n$ located in the points $x_m$ gives  
\begin{equation} \label{po}
	{\cal E}_{x_n} - {\cal E}_o  =\frac{1}{2} \sum_{n\ne m}\frac{q_nq_m}{|x_n-x_m|},
\end{equation}
which is an expression for the potential energy of a set of charges written without reference to the field they interact with. 

Finally, let us now consider the case where the location $x_n$ of the charges, or more in general the charge density $\rho(\vec x)$, is a quantum variable.  In analogy with the toy model, the system formed by the charges and the field can be described by a wave function $\Psi[A,x_n]$ which lives in the tensor product of Hilbert spaces:
\beq
| \Upsilon_{EM+S} \rangle  \in \mathcal{H}_{EM} \otimes (\otimes_n \mathcal{H}_n)
\eeq
where $\mathcal{H}_{EM}$ is the quantum space state of the field that we have described so far, and $\mathcal{H}_n$ is the Hilbert space of the $n$-th particle. 

The state of minimal energy of the field if the charges are in an (approximate) stationary eigenstate of the positions, say with value $ x_n$, is 
\begin{equation}
 \Psi_{x_n}[A]= C 
 \exp\left\{-\frac{1}{\hbar}\int d^3  x d^3  y \ \frac{B(\vec x)B(\vec y)}{|\vec x-\vec y|^2} -\frac{i}{\hbar}  \int d^3  x E_{x_n}(\vec x) A(\vec x)\right\}.
\end{equation}
We can finally write the full state that represents the field when the particles are in a quantum superposition of states with different positions $x_n^{i}$ ($n$ labels the particle and $i$ labels the branch of the wave function)\footnote{The delta function shall be understood as an approximation within the experimental resolution. We need to ensure the quasi-static condition, hence  the  momentum uncertainty of the source cannot be unbounded. }:
\begin{equation}
 \Upsilon_{x_n^{i}}[A]= \sum_i C_i
 \prod_n \delta(x_n-x_n^{i})
 \exp\left\{-\frac{1}{\hbar}\int d^3 x d^3 y \ \frac{B(\vec x)B(\vec y)}{|\vec x-\vec y|^2} -\frac{i}{\hbar}  \int d^3 x\, E_{x_n^{i}}(\vec x) A(\vec x)\right\}.
 \label{This}
\end{equation}

More in general,  consider the charge density operator $\hat{J}_0(\vec x)$ and its eigenstates  $
\hat{J}_0(\vec x)  | \phi_{\rho}\rangle =  \rho_i(\vec x) |\phi_{\rho}\rangle$. 
Consider states that are linear combinations of elements in this basis 
\beq
| \Phi_{S}\rangle =\sum_i \alpha_i  |\phi_{\rho_i}\rangle.
\eeq
Then we expect that if these states can be considered stationary, the electromagnetic field settles to its minimum energy state and the state of the full system is
\beq \label{fullEM}
 |\Upsilon_{EM+S}\rangle = \eta  \sum_i \alpha_i  |\phi_{\rho_i}\rangle |\Psi_{\rho_i}\rangle .
\eeq
In other words, the quantum state of the electromagnetic field is entangled with the source current through the information about the charge density, as  in the toy model.

The main difference between the toy model and the electromagnetic case is that in the electromagnetic case the relation between the source and the field is implemented by the Gauss constraint. In both cases the effect of a static source on a minimal energy state is the displacement of the solution by the $T$ operator.   But in the electromagnetic case, the $T$ operator shifts the state adding a \emph{longitudinal} component to the field. The quantum state of the field changes accordingly: by adding a phase in the $\Psi[A]$ representation, and by shifting support in the $\Psi[E]$ representation.  We discuss the inner product in Appendix~\ref{App:InnerProduct}.

\subsection{Field mediated Entanglement}

In a prototypical FME experiment~\cite{Bose2017}, two particles with positions $x$ and $y$ are both set into a superposition or two positions, say $x^\pm$ and $y^\pm$. The full state of the system can be written in the basis $\Psi[A,x,y]$. Before the splitting, the field is in a minimal energy state and the particles are in position eigenstates. When the particles are split in a quantum superposition, their movement excites the field away from the minimum energy state. The excitation  propagates away in the form of radiation,  dissipating energy. The emission of radiation leads the field to settle locally to a new, different, minimum energy configuration, but with a quantum split source.  That is, the total state of the particles and the field is entangled, and the state of the field takes the form in Eq.~\eqref{This} in each amplitude, with $x_n^i=\{(x^\pm,y^\pm)\}$, where $n=1,2$ and $i=\pm\pm$. In each of the four branches the expectation value of the field is different and its energy is different:
\beq
{\cal E}_{x^\pm,y^\pm} - {\cal E}_o  = \frac{q^2}{|x^\pm-y^\pm|}.
\eeq
In particular, if the distance between the two particles in one amplitude, say $|x^+-y^+|=d$, is much smaller than the distance between the particles in the other amplitudes, the gravitational energy of Eq.~\eqref{po} is
\beq
\Delta{\cal E}=\frac{q^2}{d}.
\eeq
which in a time $t$ gives rise to a relative phase 
\beq
\phi=\frac{\Delta Et}{\hbar}= \frac{q^2 t}{\hbar d}.
\eeq
This is the characteristic phase that gives rise to the entanglement that can be measured in the the FME experiments  \cite{Bose2017}. We therefore see very explicitly here that in a field description the existence of this relative phase is the effect {of the field being} in a quantum superposition of different macroscopic configurations, each with a different energy. \vfil

\section{Linearized quantum gravity in the Schr{\"o}dinger representation }
\label{sec:LinQG}

\subsection{The linearized Hamiltonian and the constraints}

 Let us now come to gravity. We use a canonical formulation, in which the metric is cast in the 3+1 decomposition~\cite{Arnowitt:1962hi}, namely
 \begin{equation}
 	ds^2 = - N^2 dt^2 + \gamma_{ij} (dx^i + N^i dt) (dx^j + N^j dt),
\end{equation}
 	where $N$ is the lapse function and $N^i $ is the shift vector. On top of this, we consider linear  perturbations around a Euclidean metric
 	 \begin{equation}
 	 	\gamma_{ij} = \delta_{ij} +  h_{ij}.
 	 	\label{h}
 	\end{equation}
 	We also expand the lapse and the shift as\footnote{For clarity of the notation, the perturbation parameter $\kappa = 16 \pi G/c^4$ is absorbed into $h_{ij}$, $n$ and $n^i$.} 
\begin{equation}
N = 1 +n,\ \  N^i =0+n^i.
\end{equation}
The linearized gravity action~\cite{Carroll:2004st}, including the coupling to the energy momentum tensor $T^{\mu\nu}$ of the matter field, reads 
	\beq\label{linearzedgravityaction}
	S = \frac{1}{4\kappa} \int  d^4x  \left(-	\partial_{\mu} {h}_{\alpha \beta} \partial^{\mu} {h}^{\alpha \beta} + 	\partial_\mu {h}  \partial^\mu {h} 	-2 \partial_{\mu} {h}^{\mu \nu} \partial_\nu {h} +2 \partial_\alpha {h}_{\mu \nu} \partial^\mu {h}^{\alpha\nu} \right) + \frac{1}{2}\int d^4x\, {h}_{\mu \nu} T^{\mu \nu} 
	\eeq
	in which $\kappa = 16 \pi G/c^4$. 
%The ADM Hamiltonian is
%\begin{equation}
 %   H_{ADM} = \int d^3x \left( N \left( - \sqrt{\gamma} R^{(3)} +\frac{1}{\sqrt{\gamma} } (\pi^{ij} \pi_{ij} - \frac{1}{2} \pi^2 ) \right) -2 N_i \sqrt{\gamma} D_j \left(\frac{\pi^{ij}}{\sqrt{\gamma}} \right) \right)
%\end{equation}
%where $D_i$ is the covariant derivative of the spatial metric, $R^{(3)}$ its the three-dimensional Ricci scalar, $\pi^{ij}$ is the canonical conjugate momentum of $\gamma_{ij}$ and $\pi = \pi^{ij} \gamma_{ij}$. More precisely, 
%\begin{equation}
%\pi^{ij} = - \sqrt{\gamma} (K^{ij} - %\gamma^{ij} K)
%\label{h}
%\end{equation}
%and $K^{ij}$ is the extrinsic curvature (or second fundamental form) of the hypersurface of constant time.  It describes the cirvature of the embedding of the spatial metric is curved in the full spacetime:
%\begin{equation}
%K_{ij} = \frac{1}{2 N} (D_i N_j +D_j %N_i -\dot{\gamma}_{ij})
%\end{equation}
Notice that to determine the time evolution to first order in our expansion we need to include in the action terms up to second order in the metric, since it is the variation of the action which determines the time evolution. The linear expansion around the Euclidean metric reduces the full diffeomorphism invariance, because a general coordinate transformation does not preserve the smallness of $h$  in Eq.~\eqref{h}.
	Under the 3+1 decomposition, this  gives us
	\beq\label{actiondecompose}
	\begin{split}
		S_{G+M}  = \frac{1}{4\kappa}   \int& d^4x  \  4n \left(   \partial_i \partial_j h^{ij}- \partial_i \partial^i h - \partial_i \partial^i n - \kappa T^{00}\right) \\
		+ & 4 n_i \left(\partial_j \dot{h}^{ij}-\partial^i \dot{h}+ \kappa T^{0i} \right)  +  \partial_j n_i  \left(  \partial^j n^i - 2\partial^i n^j\right) + 2 \kappa h_{ij} T^{ij}\\
		&\ \ \ \ + \left(\dot{h}_{ij} \dot{h}^{ij} - \dot{h}^2  - \partial_k  h_{ij} \partial^k  h^{ij}  + \partial_i h \partial^i h+ 2 \partial_k  h_{ij} \partial^j h^{ik} - 2 \partial_jh^{ij}\partial_i h \right).
	\end{split}
	\eeq
	%\LQ{I will fill in the detail on obtaining exactly (\ref{actiondecompose}) from ADM. }
 %	Due to the perturbation upon a fixed Minkowski background \fla{I don't understand the logic of this sentence, don't we do that because we want $n$ and $n_i$ to be multipliers? We would not ``need'' to do it otherwise, right?} \LQ{Yes, I thought "$n_i$ to be multiplier" should be a consequence of a physically motivated reason (for this partial gauge fixing). Perturbation upon Minkowski only fix the leading term, not this order. I need to think of a better reason.}, we need to perform a partial gauge fixing on the perturbation of Lapse and Shifts
%	 $\partial_i n = \partial_i n^j =0 $: i.e. the first order perturbation  are constant on the constant time hypersurface, then we can see immediately that $n$ and $n_i$ become Lagrangian multiplier imposing the scalar and vector constraints:

The action is independent of the time derivatives of $n$ and $n^i$. Therefore these fields behave like $A_0$ in the electromagnetic case. The vanishing of their conjugate momenta are primary constraints that give rise to secondary constraints simply obtained varying the action with respect to them. In the following, we take $\partial_i n_j =0$ and  $\partial_i n =0$. This partial gauge fixing ensures that $n$ and $n_i$ are Lagrange multipliers. The canonical momenta are
	\beq
	\pi^{kl} = \frac{1}{2\kappa} \left( \dot{h}^{kl} - \dot{h} \delta^{kl} \right), \ \  \ \ 	\{ h_{ij} (\vec{x}), \pi^{kl} (\vec{x'})\} = \delta^k_{(i} \delta^l_{j)}  \delta^3(\vec{x}-\vec{x}'),
	\eeq
where we have symmetrized on the indices $i, j$ and $k, l$.

As we did in the electromagnetic case, we focus on the situation where the sources are static or quasi static, and we set to zero all components of $T^{\mu\nu}$ except for $\kappa T^{00}\equiv\rho$.
		The constraints are the scalar constraint 
	\beq\label{constraint}
	 \mathcal{C}_{\rho} :=\partial_i \partial_j h^{ij}- \partial_i \partial^i h - \rho,
	 \eeq
	 and the vector constraint 
\beq		 
\mathcal{G}^i:=\partial_j \pi ^{ij} =0
	\eeq
They are the secondary constraints, and the commutator $\{\mathcal{C} (\vec{x}), \mathcal{G}^k (\vec{x}') \}  =0 $. 

The Hamiltonian reads\footnote{ One could obtain the same expression by the perturbative expansion of the ADM Hamiltonian~\cite{Arnowitt:1962hi} to the second order. }
	\begin{equation} \label{Hami}
	    \begin{split}
	        H_{G+M}  &= \kappa  \int d^3x  \  \left(\pi_{kl} \pi^{kl} - \pi^2/2 \right) +\\
	        &+ \frac{1}{4\kappa}   \int d^3x \left( \partial_k  h_{ij} \partial^k  h^{ij}  - \partial_i h \partial^i h - 2 \partial^k h_{ik} (\partial_j h^{ij}-\partial^i h) - 4 n_i  \mathcal{G}^i - 4n \mathcal{C}\right). 
	    \end{split}
	\end{equation}
To have  better insights of the theory, first we transform into the momentum space and split the field in its transverse and longitudinal parts:
\beq
	h^{T}_{ij}(\vec k) = P_i^kP_j^l h_{kl}(\vec k), \qquad
	h^L_{ij}(\vec k) = \Pi_i^k\Pi_j^l h_{kl}(\vec k)
\eeq
where $P^i_j(\vec k)$ and $\Pi^i_j(\vec k)$ are the projectors defined above in Eq.~\eqref{P} and Eq.~\eqref{Pi}.  This is analogous to the transverse-longitudinal decomposition in electromagnetism. The scalar constraint involves only the trace of transverse components 
		\beq
			\mathcal{C}_{\rho}  =\partial_i \partial_j h^{ij}- \partial_i \partial^i h - \rho
			 = \partial_i \partial^i  h_L- \partial_i \partial^i h - \rho  = - \partial_i \partial^i  h^T -\rho =0.	\label{Cgr}
		\eeq
	In the absence of matter, the scalar constraint ensures that $ \partial_i \partial^i  h^T =0$. In the presence of matter, it is easy to recognize it as the Poisson equation for the Newtonian potential. The vector constraint, on the other hand, depends only on longitudinal components of the canonical momenta
	\beq
	{\cal G}^i=\partial_j \pi_L^{ij} = 0.
	\label{Ggr}
	\eeq
Notice that both constraints resemble the Gauss constraint of electromagnetism, but in different ways. We will soon see how these similarities play out in the quantum theory. 

Remarkably, a detailed calculation resumed in the appendix shows that  the potential part of the Hamiltonian only depends on $h^T_{ij}$. Hence the full Hamiltonian (\ref{Hami}) is simplified to be
\beq\label{Hgr}
H = \kappa  \int \frac{d^3k}{(2 \pi)^3}  \  \left(\pi_{kl} \pi^{kl} - \pi^2/2 \right) + \frac{1}{4\kappa}   \int \frac{d^3k}{(2 \pi)^3} \left( k^2  h^T_{ij}(\vec{k})  h_T^{ij}(-\vec{k})  - k^2 h_T(\vec{k}) h_T(-\vec{k})\right)
\eeq
where we have also set the constraints to zero by choosing $n=n_i=0$,
as we did in electromagnetism.  
 At first, one may think that this is incompatible with the Newtonian field, but, as in the case of electromagnetism, this is not the case.  In fact, the field 
	\beq
	h_{ij}=\frac{2m}{r}\delta_{ij}+m t^2\partial_i\partial_j\frac{1}r =\frac{2m}{r}\delta_{ij}+\frac{mt^2}{r^3}\delta_{ij}-  \frac{mt^2}{r^5} x_ix_j
	\eeq
	is the Newtonian field of a point mass $m$ in the origin, in this gauge. It is gauge equivalent to the linearized Schwarzschild metric
	\beq
	ds^2=-\left(1-\frac{2m}r\right)dt^2+\left(1+\frac{2m}r\right)dr^2+d\Omega^2
	\eeq
	under the  change of coordinates
	\beq
	t \to t\left(1+\frac{m}{r}\right), \qquad \vec x\to \vec x\left(1+ \frac{mt^2}{r^3}\right). 
	\eeq
	
\subsection{Linearized quantum gravity in the field basis}

Here we derive the ground state wavefunctional for the linearized gravitational field in the field basis, in the case when there is a quasi-static matter source.  We fix the gauge minimally; the state is invariant under linearized spatial diffeomorphisms. The theory is defined by two constraints of Eqs.~\eqref{Cgr}-\eqref{Ggr} strongly imposed on the states and by the Hamiltonian~\eqref{Hgr} that evolves in the Minkowski coordinate time. Notice that the scalar constraint is diagonal in the configuration variables $h_{ij}$ while the vector constraint is diagonal in the conjugate momenta $\pi_{ij}$. In the following, we provide two parallel derivations of the quantum state in both bases. We check the consistency between the two descriptions by matching the results via the Fourier transform.  

\subsubsection{Solving the constraints in the metric  basis }

In the basis that  diagonalizes $h_{ij}$, we use $ \Psi [h_{ij}] : = \langle \Psi |h_{ij} \rangle $ to represent the state functional. The momentum operators act on the state as functional derivatives
\beq
\hat{\pi}^{ij}(\vec{x})=-i\hbar\frac{\delta}{\delta h_{ij}(\vec{x})}.
\eeq
The vector constraints $\mathcal{G}^i=\partial_j \pi ^{ij} $ imposed strongly on the states gives 
\beq
\partial_i \frac{\delta}{ \delta h_{ij}(\vec{x})}   \Psi [h_{ij}(\vec{x})] =0, \text{or equivalently}\ \  k_i \frac{\delta}{ \delta h_{ij} (\vec{k})}   \Psi [h_{ij}(\vec{k})] =0.
\eeq
It means that a state which is invariant under spatial diffeomorphisms is independent of the longitudinal component $h_{ij}^L$ of the field. This is analogous to the electromagnetic case, where imposing the vacuum Gauss constraint made the quantum state independent of the longitudinal component of the field.

The scalar constraint $\mathcal{C}_{\rho}    = - \partial_i \partial^i  h^T -\rho $, on the other hand, is diagonal in this basis. It restricts the support of $\Psi$ to the fields such that the trace component $h^T(\vec{x})$ satisfies the Poisson equation with source $\rho$, whose solution is  
\beq\label{Psolu}
\mathsf{h}^T_\rho(\vec {x}) =
\frac1{4\pi}\int d^3 y \frac{\rho(\vec{y})}{ |\vec{x} - \vec{y} |} + f(\vec {x}) 
\eeq
where $f(\vec{x}) $ is a harmonic function, fixed by the boundary conditions. That is, the states satisfying the constraints are given by
\beq
\Psi[h_{ij}]=\delta[\mathcal{C}_{\rho}]\ \Phi[ h_{ij}]=\delta[h^T-\mathsf{h}^T_\rho]\ \Phi[ h_{ij}].
\eeq
Since the linearized Hamiltonian is quadratic, we use the following Gaussian ansatz\footnote{Note that we do not need $h_T^2$ term in the Gaussian ansatz, because  upon imposing the scalar constraint, such term will become a constant absorbed into normalization factor.} for the ground states in the physical Hilbert space,
\beq\label{ansatzh}
	\Psi_\rho[ h_{ij}] =\delta[\mathcal{C}_{\rho}] \Phi[ h_{ij}] :=\eta \delta[\mathcal{C}_{\rho}] \exp\left\{\zeta \int \frac{d^3k}{(2\pi)^3}\  g(\vec{k})\    h^T_{ij}(\vec{k}) \  h_T^{ij} (-\vec{k})
	\right\},
\eeq
 in which $\eta$ is a normalization factor, the parameter $\zeta$ and function $g(\vec{k})$ are to be determined by solving the  time-independent Schr{\"o}dinger equation.  See Appendix~\ref{App:EigHamGrav} for details of the derivation.  We obtained $g(\vec{k}) = |k|, \zeta = - \frac{1}{4 \kappa \hbar}$. The energy eigenvalue associated to this quantum state is
\beq\label{energy}
{\cal E}_\rho  =  \hbar \int\frac{dk^3}{(2\pi)^3} |\vec{k}| \delta(0) - \frac{1}{8 \kappa} \int\frac{d^3k}{(2\pi)^3} \frac{\rho^2(\vec{k})}{|\vec{k}|^2}.
\eeq
Subtracting the vacuum energy, we obtain the self-energy of the gravitational field. In terms of the spatial coordinates, it is expressed as
\beq\label{NewtonianE}
{\cal E}_\rho - {\cal E}_{vac} = -\frac{1}{8 \kappa} \int {d^3x}\, d^3 y\  \frac{\rho(\vec x)\rho(\vec y)}{|\vec x-\vec y|}.
\eeq
In this representation, with a matter source $\rho$, the vacuum state is shifted by a $\rho$-dependent function.  The complete expression of the physical state with minimum energy is the following 
\beq\label{stateh}
	\Psi_\rho[h_{ij}]= \eta\  \delta[h^T-\mathsf{h}^T_\rho ] \exp\left\{ - \frac{1}{4 \kappa \hbar} \int \frac{d^3k}{(2\pi)^3}\  |\vec{k}|\    h^T_{ij}(\vec{k})  h_T^{ij} (-\vec{k}) 
\right\}.
\eeq
When the matter source is quasi-static, only the trace of the metric configuration changes.

\subsubsection{Solving the constraints in the momentum basis}
Now let us rederive the quantum state in  the $\pi_{ij}$ basis. We solve the constraints in the main text and give the details in Appendix~\ref{App:EigHamGrav}. In this basis, the metric operator acts on the state as a functional derivative,
\beq
\hat{h}_{ij}(\vec x)  = i \hbar  \frac{\delta}{\delta \pi_{ij} (\vec x)}.
\eeq
If we express the vector constraint  $\mathcal{G}_i =0$  in the momentum space, we immediately see that the longitudinal component of $\pi_{ij}$ is zero
\beq
k_i \pi^{ij}(\vec k) = 0 \rightarrow \pi_L^{ij}(\vec k)  = 0.
\eeq
The vector constraints are diagonal, therefore when they are imposed strongly, the quantum states which are solutions of the constraints are
\beq
\Psi[\pi] = \delta(\mathcal{G}_i) \Phi[\pi].
\eeq
Hence the solution  is constructed by the transverse component  $\pi^T_{ij}$ in momentum space. We can write a general Gaussian ansatz for the vacuum state functional which satisfies the vector constraints: 
\beq\label{ansatzpi}
\Psi_{vac}[\pi_{ij}]  =  \tilde{\eta} \exp\left(\int \frac{d^3k}{(2 \pi)^3} \tilde{g}(\vec k) \left( \alpha \pi^T_{ij}(\vec k)  \pi_T^{ij}(-\vec k) + \beta \pi_T(\vec k) \pi_T(-\vec k) \right) \right),
\eeq
where $\tilde{\eta}$ is a normalization factor, and  $ \alpha, \beta$ and the function $g(\mathbf{k})$ are yet to be determined. 

Imposing the scalar constraint strongly on the quantum state~\eqref{ansatzpi}, we obtain the condition
\beq
\hat{\mathcal{C}_{\rho}}(\mathbf{k})\Psi_\rho [\pi_{kl}] =  i \hbar k^2 P_{ij} \frac{\delta}{\delta \pi_{ij}}   \Psi_\rho [\pi_{ij}] = \rho(\vec k).
\eeq
The solution of the previous equation when $\rho(\vec k) =0$ fixes the ratio between the coefficients in Eq.~\eqref{ansatzpi},
\beq
\alpha = - 2 \beta.
\eeq
When there is a matter source, it is easy to see that the solution of the constraint has the same linear shift as in electromagnetism: 
\beq\label{rhosolution}
\Psi_{\rho}[\pi_{ij}] = \eta \exp \left( - \frac{i}{2 \hbar} \int \frac{d^3k}{(2\pi)^3} \pi_T(\vec k) \mathrm{h}_{\rho}^T(\vec k)\right) \Psi_{vac}[\pi_{ij}]
\eeq
in which $\pi_T(\vec k)$ is the trace of the transverse component of the canonical momenta, $\mathrm{h}_{\rho}^T(\vec k)$ is the Fourier mode of the solution of the Poisson equation (\ref{Psolu}), and it is a parity-even function. 
Similarly as in $h_{ij}$ basis, the constraints completely fix the vaccum state up to a momentum dependent function $\tilde{g} (\vec k)$ in the Gaussian.

We can then solve the time independent Schr{\"o}dinger equation in the $\pi_{ij}$ basis (see again Appendix~\ref{App:EigHamGrav} for details) and we finally arrive at
\beq\label{statepi}
\!\!\!\!\!\Psi_{\rho}[\pi_{ij}] = \eta \exp \left\{ - \frac{i}{2 \hbar} \int \frac{d^3k}{(2 \pi)^3} \pi_T(\vec k)  \mathrm{h}^T_{\rho}(\vec k) - \frac{\kappa} {\hbar}\int \frac{d^3k}{(2 \pi)^3} \frac{1}{|k|} \left(   \pi^T_{ij}(\vec k)   \pi_T^{ij}(-\vec k) - \frac{1}{2} \pi_T(\vec k)  \pi_T(-\vec k)  \right) \right\}.
\eeq
%\beq
%\tilde{g}(k) = \frac{1}{|k|}, \ \ \alpha = \textcolor{red}{-}  \frac{\kappa}{\hbar}
%\eeq
The energy eigenvalue is the same as in Eq.~\eqref{energy}. 

As a consistency check, we show in Appendix~\ref{App:FourierTrans} that Eq.~\eqref{stateh} and Eq.~\eqref{statepi}  are related by the Fourier transform between the two bases. We discuss the inner product in Appendix~\ref{App:InnerProduct}.

\subsubsection{The gravitational field sourced by  particles in the quantum superpositions}

When we consider a massive source in a quantum  superposition state, the structure of the quantum state is the same as in electromagnetism: the full quantum state lives on a tensor product of Hilbert spaces of the quantum source $\mathcal{H}_M$ and the gravitational field $\mathcal{H}_G$
\beq
|\Upsilon_{G+M}\rangle \in \mathcal{H}_M \otimes \mathcal{H}_G.
\eeq
Due to the scalar constraint, which is analogous to the Gauss law in linearized gravity,  the quantum state of the  field is entangled with the source through the mass density. In general, the quantum state of the quasi-static matter source could be written as superposition of mass density eigenstates:
\beq
|\Phi_M\rangle = \sum_i \alpha_i |\phi_{\rho_i} \rangle.
\eeq
Then, the full state of the gravitational field and the source is entangled, namely
\beq
|\Upsilon_{G+M}\rangle = \eta \sum_i \alpha_i  |\phi_{\rho_i} \rangle |\Psi^G_{\rho_i} \rangle.
\eeq
As the simplest example, consider the same quantum mass we have discussed in section \ref{EWaS}:
a quantum superposition of localized states with different positions $x_n^i$ ($n$ labels the particle and $i$ labels the branch of the wave function). Its  full quantum state is
\begin{equation}
    \begin{split}
        \Upsilon_{x_n^{i}}[x_n, \pi_{ij}]&= \sum_i C_i
 \prod_n \delta(x_n-x_n^{i})\cdot \\
   &\cdot \exp \left\{  - \frac{\kappa} {\hbar}\int d^3x d^3y \frac{   \pi^T_{ij}(\vec x)   \pi_T^{ij}(\vec y) - \frac{1}{2} \pi_T(\vec x)  \pi_T(\vec y)  }{{|\vec x-\vec y|^2}} - \frac{i}{2 \hbar} \int d^3x \pi_T(\vec x)  \mathrm{h}^T_{x_n^{i}}(\vec x)\right\}.
    \end{split}
\end{equation}
The relative phase can be obtained by calculating the energy difference between the different configurations from the expression for the energy given in Eq.~\eqref{NewtonianE}. The relative phase between the closest amplitudes (separated by a distance $d$), corresponding to the phase measured in the the FME experiment, is
\beq
\phi=\frac{\Delta Et}{\hbar}= -\frac{m^2 t}{8 \kappa \hbar d}.
\eeq

%Finally, for consistency check, we show in Appendix that (\ref{stateh}) and (\ref{statepi})  are precisely related by the Fourier transform between $h_{ij}$ and $\pi_{ij}$. \LQ{This is the last step that hasn't matched precisely yet. All other things are fine.}

\section{Discussion}
\label{sec:Discussion}

The state of a quantum field in the presence of charges can be written explicitly and in a straightforward manner in the field basis.  In Table~\ref{tab:table1} we summarize the similarities between the descriptions of the electromagnetic and the linearized gravity cases. When the charges are in quantum superposition of distinct density eigenstates, the minimal energy state on which the field settles after dissipating away all possible energy is an entangled state of the charges and field degrees of freedom. This formulation could also be used to describe the quantum spacetime for the gravitational switch~\cite{Zych:2017tau}, where the order of application of quantum operations is in a quantum superposition due to the source mass of the gravitational field being in a quantum state.

\begin{table}[t]\label{compare}
  \begin{center}
    \setlength\extrarowheight{7 pt}
    \begin{tabular}{p{0.21\textwidth}|p{0.33\textwidth}|p{0.28\textwidth}} 
       &\textbf{Electromagnetism} & \textbf{Linearized Gravity} \\
         \hline
      Temporal gauge & $A_0 =0$ & $h_{0\mu} = 0$  \\
     \hline
     Canonical variables & $	\left\{ A_i(\vec x), E_j(\vec x') \right\}$  & $\{ h_{ij} (\vec x), \pi^{kl} (\vec x')\} $ \\
      \hline
       No. of constraints & $1$ & $4$  \\
       \hline
      Similar constraints & Gauss law in $A$ basis & Vector constraint in $h$ basis \\
       (without matter)  &  $\partial_j \frac{\delta}{\delta A_j(\vec{x})} \Psi[A] = 0$  &  $\partial_i \frac{\delta}{ \delta h_{ij} (\vec x)}   \Psi [h_{ij}] =0$ \\
       \hline
        Similar constraints & Gauss law in $E$ basis with charge &  Scalar constraints \\
     (with matter) & $\nabla \cdot E = \Delta \phi = \rho$ &  $ \Delta h^T = -\rho $ \\
      \hline
       Vaccum state & Gaussian of transverse mode &  Gaussian of transverse mode with zero trace\\
             \hline
     The d.o.f activated with a static source &   Longitudinal mode $A_L$ &  Trace of transverse mode $h_T$ \\
    \end{tabular}
     \caption{\label{tab:table1} Comparison between the canonical descriptions of the electromagnetic and the linearized gravity. }
  \end{center}
\end{table}

It is important to remark that, although the regime of FME is captured by a perturbative description of gravity as a quantum field on a classical background, a gravitational source in a quantum superposition fundamentally realizes a quantum superposition of spacetimes. Our explicit description emphasises this aspect, by providing an explicit quantum state associated to the two different configurations of gravity (see also the arguments in~\cite{belenchia2018quantum, belenchia2019information, christodoulou2019possibility}).

A few considerations are important in view of the discussion on the interpretation of the FME experiments in the context of an effective quantum field theory of the gravitational field:

\noindent\textbf{Separation between transverse and longitudinal components of the field.} %The split of the field in different components (transverse, longitudinal, trace) does not correspond to a separation of physical degrees of freedom. First, 
It has been argued that the Coulomb/Newtonian field should not be quantized, because it is just a gauge potential fully determined by the charge/matter. In this view, the true quantum degrees of freedom are the transverse modes, i.e. photons/gravitons. Hence, the longitudinal component of the field cannot be in a quantum superposition, and experiments solely involving the Newtonian interaction can only test the quantum degrees of freedom of matter, and not of gravity. 

Our calculation explicitly shows that in linearized quantum gravity, instead, the Coulomb /Newtonian field has a quantum state entangled with the source.  The split between  the transverse and the longitudinal component of the field is not an absolute separation of degree of freedom; it is a momentum frame dependent notion. Moreover, a relativistic field, whether classical or quantum, is a local entity, with a local dynamics that does not transmit information faster than light. The split is by definition highly non-local and it cannot be distinguished by local observations in the laboratory. Let us consider the following example. When moving a charge with an external force, the longitudinal component of its field changes instantaneously all over the universe. But of course the field cannot change at a distance $r$ before a time $t=r/c$ has elapsed.   This clearly shows that the "Coulomb field" or the "Newtonian field" are  mathematical artefacts that may be useful at small distances, but cannot be used to deduce any physical property. These aspects are also discussed in Ref.~\cite{Christodoulou:2022knr}.

%These considerations are relevant for the discussion on gravity experiments, because it has been argued that only the transverse field is quantized. According to this view, the longitudinal component of the field cannot be in a quantum superposition, and experiments solely involving the Newtonian interaction can only test the quantum degrees of freedom of matter, and not of gravity. 

As clearly demonstrated in this work, in a low-energy description of gravity as a quantum field, the statement that only the transverse modes can be quantized is incorrect: in the presence of quantum matter, the gauge-invariant quantum wave functional contains the longitudinal components through an integration of all momentum frames. It can be entangled with matter and be in a superposition.
%The field is a single entity, that cannot be split.  
%If relativistic quantum field theory is correct, a quantum split source leads the field to settle on a state that is itself in quantum superposition.  
Such superposition of classical macroscopic fields is cumbersome to be represented in the Fock basis, but it is very natural and transparent to be described in the (mathematically equivalent) field basis. 

\noindent\textbf{Hilbert space factorization.} It has been argued that in the FME experiment the particles cannot be initially prepared in a product state, due to the gauge constraints and long-range nature of the Newtonian potential. 
As we have shown, when we consider the quantization of the gravitational field with a quantum source that is externally controlled, the Hilbert space of the source and the field factorizes into a tensor product\footnote{Note that we are studying a much simpler situation than the full-fledged interacting quantum field theory. We neglect the possibility of backreaction from the field to the motion of the source and the creation/annihilation of particles.}. What cannot be factorized is the Hilbert space of the gravitational field in different regions in spacetime.

\noindent \textbf{Dynamical transient between different energy configurations.} The entanglement we have  illustrated assumes that the field relaxes to its minimal energy state after the charges are modified.  This requires the field to radiate away the excitations  generated by the displacement of the charges.   This never happens completely, of course, but it happens with a good approximation in a compact region, after a time of the order of the light transit time in the region (assuming no mirrors around). It is only after this time that in this compact region the field is approximated again by a minimal energy states---this time the new minimal energy state determined by the new position of the charges. 

The detection of charges-field entanglement and its effects is not yet achievable with current experimental technologies. For this reason, it is important to look for other novel effects potentially observable on a shorter timescale. The results in our paper not only provide a theoretical foundation of the FME experiment, but will also  be useful for developing new proposals to test quantum features of gravity in table-top experiments.

%Measuring these effects for the electromagnetic field would be extremely interesting, and doing so for gravity would be a major step towards quantum tests of the nature of gravity. 

\acknowledgments{We want to thank Eugenio Bianchi, Glenn Barnich and Ognyan Oreshkov for discussions.
This work is partly funded by the ID\# 61466 grant from the John Templeton Foundation, as part of the ``The Quantum Information Structure of Spacetime (QISS)'' Project (\href{qiss.fr}{qiss.fr}). F.G. acknowledges support from the Swiss National Science Foundation via the Ambizione Grant PZ00P2-208885. F.G. and C.R. acknowledge support from Perimeter Institute for Theoretical Physics. Research at Perimeter Institute is supported in part by the Government of Canada through the Department of Innovation, Science and Economic Development and by the Province of Ontario through the Ministry of Colleges and Universities.}

\appendix
\section{Commutator of the constraints}
Here we show that the vector constraint and scalar constraints are indeed commute:
	\beq \begin{split}
		\{\mathcal{C} (\vec x), \mathcal{G}^k (\vec x') \} & = \int d^3y \frac{\delta \left(\partial_i \partial_j h^{ij}(\vec x )- \partial_i \partial^i h (\vec x ) \right)}{ \delta h_{mn} (\vec y )} \frac{\delta \partial_\beta \pi^{\beta k} (\vec x' )}{ \delta \pi^{mn} (\vec y )}  \\
		& =  \int d^3y \left( \partial^m \partial^k_{(x)}  - \delta^{mk}  \partial_i \partial^i_{(x)}   \right) \delta^3(\vec x -\vec y ) \partial_m^{(x')} \delta^3(\vec x' -\vec y ) \\
		& =   -\left( \partial_m^{(x')}  \partial^m_{(x)} \partial^k_{(x)}  -  \partial^k_{(x')}  \partial_i \partial^i_{(x)}   \right) \delta^3(\vec x -\vec x' ) \\ 
		& =  -  \partial_m^{(x)}  \partial^m_{(x)} \partial^k_{(x')} \delta^3(\vec x -\vec x' )  + \partial^k_{(x')}  \partial_i \partial^i_{(x)}    \delta^3(\vec x -\vec x' ) =0 
	\end{split}
	\eeq 
	They are both first class secondary constraints. 
\section{Modes decomposition and simplification  of the linearized gravity hamiltonian}
\label{App:LinGravHam}
In the temporal gauge $n = n_i =0$, the full Hamiltonian (\ref{Hami})  expressed in the momentum space is the following
\beq
\begin{split}
H &=   \kappa  \int \frac{d^3k}{(2 \pi)^3}  \  \left(\pi_{ij}(\vec k) \pi^{ij}(-\vec k)+ \frac{\pi(\vec k) \pi(-\vec k)}{2} \right) +\\
&+ \frac{1}{4\kappa}   \int \frac{d^3k}{(2 \pi)^3} \left( \underbrace{k^2  h_{ij}(\vec k)  h^{ij}(-\vec k)  - k^2 h(\vec k) h(-\vec k)}_{A} - \underbrace{2 k^l h_{il}(\vec k) (k_j h^{ij}(-\vec k) -k^i h(-\vec k))}_{B} \right).
\end{split}
\eeq
It turns out that the potential part of the Hamiltonian is independent of the longitudinal components  and can be further simplified. Under mode decomposition, there are three types of terms: the transverse terms, the longitudinal terms and the mixture of  longitudinal and transverse terms. Remarkably, the latter two types of terms in A and B  cancel:  $A_{LL} - B_{LL} = 0$, $A_{LT} - B_{LT} =0$. More precisely:
\beq
\begin{split}
& A_{LL} =B_{LL} = \int\frac{d^3 k}{(2 \pi)^3} \left( 2 k^l k_j  h_{il}(\vec k) h^{ij}(\vec k)  - \frac{2}{k^2} k_m k_n h^{mn}(\vec k) h^{ij}(\vec k)  k_i k_j \right)\\
&A_{LT} =B_{LT} = - 2 \int\frac{d^3 k}{(2 \pi)^3}  P^{ij} h_{ij}(\vec k)h_{mn}(\vec k) k^m k^n.
\end{split}
\eeq
Hence, the potential only depends on the transverse terms, and the hamiltonian simplifies to
\beq
H = \kappa  \int \frac{d^3k}{(2 \pi)^3}  \  \left(\pi_{kl} \pi^{kl} - \pi^2/2 \right) + \frac{1}{4\kappa}   \int \frac{d^3k}{(2 \pi)^3} \left( k^2  h^T_{ij}(\vec k)  h_T^{ij}(-\vec k)  - k^2 h_T(\vec k) h_T(-\vec k)\right). 
\eeq

\section{Solving the ground state of the gravitational field in both representations}
\label{App:EigHamGrav}

In the $h_{ij}$ basis,  the kinetic part of the hamiltonian operator is represented as
\beq
\hat{T}  =   \kappa\ \hbar^2 \int \frac{d^3k }{(2 \pi)^3}  \left(-  \frac{\delta}{\delta h_{ij}(\vec k)}\frac{\delta}{\delta h^{ij}(-\vec k)  } +\frac{1}{2}\delta_{mn} \frac{\delta}{\delta h_{mn}(\vec k)} \delta_{kl}\frac{\delta}{\delta h_{kl}(-\vec k)}  \right)
\eeq
Since the constraints are first class, we can first solve the time-independent Schr{\"o}dinger equation and find the eigenstates of the hamiltonian, and then impose the scalar constraint.
The first derivative acting on the Gaussian ansatz of Eq.~\eqref{ansatzh} gives
\beq
\frac{\delta}{\delta h^{ij}(-\vec k)}\Phi [h_{ij}] = 2 \zeta g(- \vec k)  P_i^{m} P_j^{n} h_{mn}(\vec k).
\eeq
After taking two functional derivatives, we obtain the eigenfunction of the kinematic energy operator
\beq
\hat{T} \Phi [h_{ij}] = 2 \kappa \hbar^2 \zeta^2 g^2(\vec k) \left(h_T(\vec k) h^T(-\vec k))  - 2  h_T^{ij} (\vec k) h^T_{ij}(-\vec k) \right) \Phi_{\rho} [h] - 4 \kappa \zeta \hbar^2  g(- \vec k) \Phi_{\rho} [h] 
\eeq
To ensure that $\Psi_{\rho}[h_{ij}] = \delta[\mathcal{C}] \Phi [h_{ij}] $ is the ground state of the Hamiltonian, upon imposing the scalar constraint, the eigenfunction of the kinematic operator needs to cancel  the potential part of the Hamiltonian.  This  gives us
\beq
\zeta = - \frac{1}{4 \kappa \hbar}, \ \ g(\vec k) = |\vec k|
\eeq
in which the minus sign in $\zeta$ ensures that the ground state has positive vacuum energy. The eigenvalue of the Hamiltonian is
\beq
{\cal E}_\rho  =  \hbar \int\frac{dk^3}{(2\pi)^3} |\vec{k}| \delta(0) - \frac{1}{8 \kappa} \int\frac{d^3k}{(2\pi)^3} \frac{\rho^2(\vec{k})}{k^2}.
\eeq
Now let us rederive the eigenstates in the $\pi_{ij}$ basis. The operators $\hat{h}^T_{ij} (\vec k)$ and $\hat{h}^T (\vec k)$ are expressed as functional derivatives\footnote{We want to emphasize a subtlety of the derivation, namely that the transverse component of the canonical momenta, $\pi_{ij}^T = P_i^k P_j^l \pi_{kl}$, is not the canonical conjugate of the transverse metric perturbation $h_{ij}^T = P_i^k P_j^l h_{kl}$:
\beq
\hat{h}_{ij}^T (\vec k) \neq i \hbar \frac{\delta}{\delta \pi_T^{ij} (\vec k)}, \ \ \delta \pi^T_{kl} (\vec k) \neq  P^i_k P^j_l  \delta \pi^T_{ij} (\vec k).
\eeq
This is easy to verify classically:
\beq
\pi_{ij}^T (\vec k) = P_i^k P_j^l \pi_{kl}(\vec k) \neq \frac{\delta \mathcal{L}}{\delta \dot{h}_T^{ij}}.
\eeq
}

 \beq
\hat{h}_{ij}^T (\vec k)  = i \hbar P^k_i P^l_j \frac{\delta}{\delta \pi_{kl}(\vec k)} , \ \  \hat{h}^T (\vec k)  = i \hbar P^{kl}  \frac{\delta}{\delta \pi_{kl}(\vec k)}
\eeq
As we have shown in the main text, in the $\pi_{ij}$ representation, the solution to both scalar and vector constraints are in Eq.~\eqref{ansatzpi} and Eq.~\eqref{rhosolution}. Now we can solve the Schr{\"o}dinger equation to determine the coefficients in the Ansatz.  

The kinematical term of the Hamiltonian is diagonal in this basis, and the potential term reads
 \beq 
\!\!\!\!\!\!\!\!\hat{V} = - \frac{\hbar^2}{4\kappa}   \int \frac{d^3k}{(2 \pi)^3} k^2 \left(  P^k_i(\vec k) P^l_j(\vec k) \frac{\delta}{\delta \pi_{kl}(\vec k)}  P_m^i(-\vec k) P_n^j(-\vec k) \frac{\delta}{\delta \pi_{mn}(-\vec k)}  -  P^{kl}(\vec k)  \frac{\delta}{\delta \pi^{kl}(\vec k)}  P^{ij}(-\vec k) \frac{\delta}{\delta \pi^{ij}(-\vec k)} \right).
\eeq
The first functional derivative acting on Eq.~\eqref{rhosolution} gives us
\beq
\frac{\delta}{\delta \pi_{ij}(-\vec k)} \Psi_{\rho} [\pi]  = \left( 2 \tilde{g}(-\vec k) \left( \alpha \pi_T^{ij}(\vec k) + \beta P^{ij} \pi^T(\vec k) \right) - \frac{i}{2 \hbar} P^{ij} \mathrm{h}_{\rho}^T(\vec k) \right) \Psi_{\rho} [\pi]. 
\eeq
Hence  we obtain
\beq\label{VPsi}
\begin{split}
\hat{V} \Psi_{\rho} [\pi] & = - \frac{\hbar^2}{4\kappa}   \int \frac{d^3k}{(2 \pi)^3} k^2   P^k_i P^l_j \frac{\delta}{\delta \pi_{kl}(\vec k)}  P_m^i P_n^j \frac{\delta}{\delta \pi_{mn}(-\vec k)}   \Psi_{\rho} [\pi]\\
& = -\frac{1}{\kappa} \int \frac{d^3k}{(2 \pi)^3} k^2  \left(   4 \hbar^2  \beta^2 \tilde{g}(\vec k)^2 ( \pi_T^{ij} \pi^T_{ij} - \frac{1}{2} \pi_T^2 )- 2\hbar^2 \beta \delta(0) \tilde{g}(-\vec k) + \frac{1}{8} \mathrm{h}_{\rho}^T(\vec k)^2 \right)  \Psi_{\rho} [\pi]
\end{split}
\eeq
Upon imposing the vector constraint $\pi^L_{ij} =0$, 
the kinetic part of the hamiltonian becomes 
\beq
T\overset{\mathcal{G}_i}{=} \kappa  \int \frac{d^3k}{(2 \pi)^3}  \  \left(\pi^T_{kl} \pi_T^{kl} - \pi_T^2/2 \right).
\eeq
Comparing the previous equation with the first term in the second line of Eq.~\eqref{VPsi}, we obtain 
\beq
g(\vec k) = |\vec k|, \ \ \ \beta = \frac{\kappa}{2 \hbar}.
\eeq
With this, we finally find the ground state in Eq.~\eqref{statepi}. From Eq.~\eqref{VPsi} we calculate the energy eigenvalues, which coincide with those in Eq.~\eqref{energy}.

\section{The inner product of the states}

\label{App:InnerProduct}

In this section, we calculate the inner product between different quantum states and determine the normalization factor.

Let us start from the EM vacuum state of Eq.~\eqref{eq:VacuumEMFree}. As the state only depends on the transverse mode, we can separate the integration measure as
\beq
\begin{split}
\int \mathcal{D}[A] \Psi_0 [A] \Psi^*_0 [A] &= C^2   \int \mathcal{D}[A_L] \mathcal{D}[A_T]  \exp\left\{-\frac{1}{\hbar}\int \frac{d^3 k}{(2\pi)^3} |\vec{k}| \left(\tilde{A}_i(\vec{k})  P^i_j(\vec{k}) 
\tilde{A}^j(-\vec{k}) \right)\right\}\\
&=C^2  \prod_k \int d A_L(\vec{k}) dA_T(\vec{k})  \exp\left\{-\frac{1}{\hbar} \frac{|\vec{k}|}{(2\pi)^3}  \left(\tilde{A}_i(\vec{k})  P^i_j(\vec{k}) 
\tilde{A}^j(-\vec{k}) \right)\right\} \\
& = C^2 \prod_k  2\pi^2 \sqrt{\frac{2 \hbar}{|\vec{k}|}}  \int d A_L(\vec{k}) \equiv 1
\end{split}
\eeq
in which the functional integral can be written as  $\int \mathcal{D} A := \prod_k \int d A(k)$, and we have performed the Gaussian integral of $A_T$ for each momenta. Therefore, the normalization factor is
\beq
C^2 = V_{gauge}^{-1}  \prod_k 2^{-3/2} \pi^{-2} \sqrt{\frac{\omega_k}{  \hbar}},
\eeq
This is almost the same as the normalization of a scalar field vacuum state, 
up to  $V_{gauge} =\prod_k\int   d A_L(\vec{k}) $ which measures the volume of the gauge orbit.

For a stationary charge, when the electromagnetic field settles to its minimum energy state (new vacuum), the full quantum state is formally written as in Eq.~\eqref{fullEM}: 
\beq
|\Upsilon_{EM+S}\rangle = \eta_{EM}  \sum_i \alpha_i  |\phi_{\rho_i}\rangle |\Psi_{\rho_i}\rangle .
\eeq
in which $| \phi_{\rho_i}\rangle$ is the eigenstate of the charge density operator (e.g. a semi-classical state of the charge maximally localized in position and momentum). In each amplitude of the superposition, the linear shift of the phase cancels out when we evaluate the integral. We thus find the normalization coefficient 
\beq
\eta_{EM}^2 = C^2_{vac} /\sum_i |\alpha_i|^2.
\eeq

For the quantum state of linearized gravity, it is more convenient to evaluate the inner product in the $\pi_{ij}$ representation of Eq.~\eqref{statepi},
\beq
\begin{split}
\int \mathcal{D}[\pi] \Psi_\rho [\pi_{ij}] \Psi^*_\rho [\pi_{ij}] & = \eta^2 \int \mathcal{D} \pi^T_{ij} \mathcal{D} \pi^L_{ij}  \exp \left\{ - \frac{2 \kappa} {\hbar}\int \frac{d^3k}{(2 \pi)^3} \frac{1}{|\vec{k}|} \left(   \pi^T_{ij}(\vec k)   \pi_T^{ij}(-\vec k) - \frac{1}{2} \pi_T(\vec k)  \pi_T(-\vec k)  \right) \right\}\\
&=  \eta^2 \prod_k \int d \pi^L_{ij}(\vec{k}) d\tilde{\pi}^T_{ij}(\vec{k}) d\pi^T(\vec{k})  \exp \left\{ - \frac{2 \kappa} {(2 \pi)^3 \hbar |\vec{k}| }   \tilde{ \pi}^T_{ij}(\vec k)   \tilde{\pi}_T^{ij}(-\vec k)  \right\}\\
&=  \eta^2 \prod_k 2 \pi^2 \sqrt{\frac{ \hbar \omega_k }{ \kappa}} \int d \pi^L_{ij}(\vec{k})  d\pi^T(\vec{k})  \equiv 1,
\end{split}
\eeq
in which we have redefined the traceless variable as $\tilde{\pi}^T_{ij}(\vec{k}) = \pi^T_{ij}(\vec k) -\frac{P_{ij}(\vec k)}{2} \pi^T (\vec k)$ and the integration measure splits as
$\mathcal{D} \pi^T_{ij} = \mathcal{D} \tilde{\pi}^T_{ij} \mathcal{D} \pi^T$. Analogously to the EM case, this procedure fixes the normalization factor $\eta$.

\section{Consistency check: transformation between $h_{ij}$ and $\pi_{ij}$ representations}
\label{App:FourierTrans}

When we change the basis between $h_{ij}$ and $\pi_{ij}$, we perform a functional Fourier transform, defined as
\beq
\begin{split}
 F[h_{ij}] &=\int \mathcal{D} \pi_{ij} \exp\left\{\frac{i}{\hbar} \int \frac{d^3k}{(2 \pi)^3} \pi^{kl}(\vec k) h_{kl}(-\vec k) \right\}  \tilde F[\pi_{ij}] \\
 &= \int \mathcal{D} \pi^T_{ij} \mathcal{D} \pi^L_{ij} \exp\left\{\frac{i}{\hbar} \int \frac{d^3k}{(2 \pi)^3} \left(\pi_T^{kl}(\vec k) h^T_{kl}(-\vec k) + \pi_L^{kl}(\vec k) h^L_{kl} (-\vec k)\right) \right\} \tilde F[\pi_{ij}] 
 \end{split}
\eeq
We start from the quantum state of Eq.~\eqref{statepi} in the $\pi_{ij}$ basis. As the quantum state does not depend on the longitudinal component  $\pi^L_{ij} $, this is simply integrated as a delta function. The Fourier transform of the quantum state then is
\beq\label{Fourier}
\begin{split}
\Psi_{\rho}[h_{ij}]:=\int \mathcal{D} \pi^T_{ij} \exp &\left \{\frac{i}{\hbar} \int \frac{d^3k}{(2 \pi)^3} \left(\pi_T^{kl}(\vec k) h^T_{kl}(-\vec k)  - \frac{1}{2} \pi_T(\vec k)  \mathrm{h}^T_{\rho}(\vec k)\right)\right\}\cdot\\
&\cdot \exp \left\{- \frac{\kappa} {\hbar}\int \frac{d^3k}{(2 \pi)^3} \frac{1}{|\vec k|} \left(   \pi^T_{ij}(\vec k)   \pi_T^{ij}(-\vec k) - \frac{1}{2} \pi_T(\vec k)  \pi_T(-\vec k)  \right) \right\}
\end{split}
\eeq
By completing the square, the argument of the exponential function can be rewritten as
\begin{equation}
   \begin{split}
        &-\frac{\kappa}{\hbar}\int \frac{d^3 k}{(2\pi)^3}\frac{1}{|\vec{k}|}\left( \tilde{\pi}^T_{ij}  -i\frac{|\vec{k}|}{2\hbar}h^T_{ij}(\vec{k})\right)\left( \tilde{\pi}_T^{ij}(-\vec k)  -i\frac{|\vec{k}|}{2\hbar}h_T^{ij}(-\vec{k})\right)+\\ &+\frac{i}{2\hbar}\int\frac{d^3 k}{(2\pi)^3}\left( h_T(\vec{k}) - h_\rho^T(\vec{k}) \right)\pi^T(\vec{k})- \frac{1}{4\hbar\kappa}\int\frac{d^3 k}{(2\pi)^3}|\vec{k}|h_{ij}^T(\vec{k})h^{ij}_T(-\vec{k})
   \end{split}
\end{equation}
where we redefine a traceless variable $\tilde{\pi}^T_{ij}(\vec{k}) = \pi^T_{ij}(\vec k) -\frac{P_{ij}(\vec k)}{2} \pi^T (\vec k)$.

%\beq
%- \left(\sqrt{\frac{\kappa}{|k|}}(\pi^T_{ij}(\vec k) -\frac{P_{ij}(\vec k)}{2} \pi^T (\vec k)) + i ( x h^T_{ij}(\vec k) + y \frac{P_{ij}(\vec k)}{2} h^T(\vec k) ) \right)^2
%\eeq
%by matching with the phase in Eq.~\eqref{Fourier}, we obtain
%\beq
%x = - \frac{1}{2} \sqrt{\frac{|\vec k|} {\kappa}}
%\eeq
%while y is completely unconstrainted. Upon imposing scalar constraint, $h^Th_T$ type of term in the Gaussian will only change the normalization factor of the whole state, hence we can simply set $y=0$ (or an arbitrary number).  Putting everything together, we have
Notice that we could have completed the square by adding a term proportional to $h_T(\vec{k})$ in the square brakets in the first line of the last equation. However, upon imposing the scalar constraint, this term would just amount to a constant that could be absorbed in the normalisation factor. Hence, the state above is fully general.

The Fourier-transformed state then is
\beq \label{eqapp:FourierInt}
\begin{split}
\Psi_{\rho}[h_{ij}]&:=\int \mathcal{D} \pi^T_{ij} \exp\left\{-\frac{\kappa}{\hbar}\int \frac{d^3 k}{(2\pi)^3}\frac{1}{|\vec{k}|}\left( \tilde{\pi}^T_{ij}  -i\frac{|\vec{k}|}{2\hbar}h^T_{ij}(\vec{k})\right)\left( \tilde{\pi}_T^{ij}(-\vec k)  -i\frac{|\vec{k}|}{2\hbar}h_T^{ij}(-\vec{k})\right) \right\} \cdot\\
&\cdot \exp\left\{\frac{i}{2\hbar}\int\frac{d^3 k}{(2\pi)^3}\left( h_T(\vec{k}) - h_\rho^T(\vec{k}) \right)\pi^T(\vec{k})- \frac{1}{4\hbar\kappa}\int\frac{d^3 k}{(2\pi)^3}|\vec{k}|h_{ij}^T(\vec{k})h^{ij}_T(-\vec{k}) \right\}.
\end{split}
\eeq
We now notice then, upon changing the functional integration variable to $\mathcal{D} \pi^T_{ij} = \mathcal{D} \tilde{\pi}^T_{ij} \mathcal{D} \pi^T$, we can solve the transverse traceless part $\tilde{\pi}^T_{ij}$ as a Gaussian functional integral, and the trace part $\pi^T$ as a delta function. The final result in the $h_{ij}$ basis is
\beq
\Psi_{\rho}[h_{ij}]= \eta \delta[h^T- \mathrm{h}^T_{\rho} ] \exp\left\{-\frac{1}{ 4 \kappa \hbar} \int \frac{d^3k}{(2 \pi)^3} |\vec k| h^T_{ij} (\vec k) h_T^{ij} (-\vec k)\right\}.
\eeq
which coincides with Eq.~\eqref{stateh}.

\bibliographystyle{unsrturl}
\bibliography{main.bib}

\end{document}